\documentclass[aps,prd,letterpaper,11pt,twoside,tightenlines,nofootinbib,showpacs,preprint,onecolumn]{revtex4}
\usepackage{graphicx}
\usepackage[sort&compress]{natbib}
\usepackage{latexsym}
\usepackage{epsfig}
\usepackage[english]{babel}
\usepackage{graphicx}
\usepackage[T1]{fontenc}
\usepackage[utf8]{inputenc}
\usepackage{amsmath}

\usepackage[figurename=FIG.,tablename=TABLE,font=footnotesize]{caption}
\usepackage{setspace}

\usepackage{bm}

\begin{document}
\newcommand{\eg}{{\it e.g.}}
\newcommand{\etal}{{\it et. al.}}
\newcommand{\ie}{{\it i.e.}}
\newcommand{\be}{\begin{equation}}
\newcommand{\dd}{\displaystyle}
\newcommand{\ee}{\end{equation}}
\newcommand{\bea}{\begin{eqnarray}}
\newcommand{\eea}{\end{eqnarray}}
\newcommand{\bef}{\begin{figure}}
\newcommand{\eef}{\end{figure}}
\newcommand{\bce}{\begin{center}}
\newcommand{\ece}{\end{center}}

\centerline{\bf Quantitative predictions of neoadjuvant chemotherapy effects in breast cancer}
\vskip10 pt
\centerline {\bf by individual patient data assimililation}
\vskip30pt
\centerline{P. Castorina $^{(a,b,*)}$, D.Carco'$^{(a)}$, C.Colarossi$^{(a)}$, M.Mare$^{(a,e)}$, L.Memeo$^{(a)}$},
\vskip 10pt
\centerline{M.Pace$^{(b,c,d)}$, I.Puliafito$^{(a)}$, D.Giuffrida$^{(a)}$}
\vskip 20pt
\centerline{${}^a$ Dipartimento di Oncologia Sperimentale,Istituto Oncologico del Mediterraneo , 95029 Viagrande, Italy}
\vskip 20pt
\centerline{${}^b$ INFN, Sezione di Catania, I-95123 Catania, Italy.}
\vskip 10pt
\centerline{${}^c$ Scuola di Specializzazione in Fisica Medica, Universita' di Catania, Italy}
\vskip 10pt
\centerline{${}^d$ Centro Siciliano Fisica Nucleare e Struttura della Materia, Catania, Italy}
\vskip 10pt
\centerline{${}^e$ Dipartimento di Scienze Biomediche,odontoiatriche}
\centerline{e delle immagini morfologiche e funzionali, Universita' di Messina,Italy}

\vskip 50pt
(*) corresponding author, paolo.castorina@ct.infn.it

\vfill
\eject


\vskip 20pt

{\bf Abstract}

\vskip 30 pt

Neoadjuvant chemotherapy has been used for  breast cancer aiming at downgrading before surgery. In this article we propose
a new quantitative analysis of the effects of the neoadjuvant therapy to obtain numerical, personalized, predictions on the shrinkage of the tumor size after the drug doses, by data assimilation of the  individual patient. 
The algorithm has been validated by a sample of 37 patients with histological diagnosis of locally advanced primary breast carcinoma. The biopsy specimen, the initial tumor size and  its reduction after each treatment were known for all patients.
We find that: a) the measure of tumor size at the diagnosis and after the first dose permits to predict the size reduction for the follow up; b) the results are in agreement with our data sample, within 10-20 $\%$, for about 90$\%$ of the patients.
The quantitative indications suggest the best time for surgery. The analysis is patient oriented, weakly model dependent and can be applied to other cancer phenotypes.

\vfill
\eject

\section{Introduction}

In the era of personalized oncology, mathematical models are a useful tool for a better understanding of the clinical effects of  therapy. 

The expected  individual response to tumor therapies is generally based on a set of indices, defined  with large quantitative variability.  
For example, for neoadjuvant chemotherapy for locally advanced breast cancer, aiming at downgrading before surgery,  one usually considers the subtypes classification according to
the expression of hormone receptors, estrogen (ER) and progesteron(PR), of the human epidermal growth factor receptor 2 (HER2), the proliferation index ki67, the inizial tumor size and cellularity. Indeed, these clinical informations may have a prognosis value similar to that of multigene prognostic score \cite{cuzik}.

The tumor progression during neoadjuvant chemotherapy \cite{claude,fukuda}, described by previous ( and others)  predictive factors,  gives
direct informations on the response to the  therapy. By those analyses one gets  semi-quantitative results following the standard classification:  tumor size (median and range), T stage, Node stage, JACC stage, Lymphovascular invasion and other parameters.

A complementary strategy could be obtained by more quantitative informations , based on 
 numerical approaches which, by  single patient data assimilation, enhance the  level of reliability  of forecasts on the individual response.

Here we discuss an algorithm which, starting from the measure of the tumor size ( radius ) at the diagnosis and  after the first dose, is able to predict, essentially without free parameters,
the shrinkage of the tumor in the sequence of treatments. The proposed method is , by itself, patient oriented since the first size reduction  and the initial cellularity take into account the specific initial  condition. 

The numerical predictions agree , within $10-20\%$, for more than $90 \%$ of the observed data of our sample of 37 patients.

\begin{figure}
\epsfig{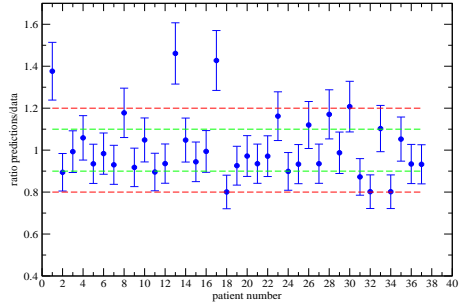}
\caption{Predictions to data ratio on the  tumor radius after the second dose. The green (red)  line indicates the $1 \pm 0.1$ ($1 \pm 0.2$) interval.}
\end{figure}

\section{Mathematical formulation of the diagnostic algorithm}

The breast tumor growth is described by the Gompertz law \cite{gompertz,norton1}, solution of the differential equation
\be
({1}/{N}) dN/dt = - k  ln ({N}/{N_\infty}),
\ee
where $N$ is the cell number at time $t$, $k$ is a constant and $N_\infty$ is the maximum number of cells ($N_\infty = 3.1 *10^{12}$ , according to ref. \cite{norton1}).

The modification of the specific growth rate due to  chemotherapy, during the time interval of a single treatment, is obtained  by  introducing  a function $c(t)$  in the previous equation \cite{norton2,norton3,norton4,castorina}, i.e. 
\be
({1}/{N}) dN/dt = - k  ln ({N}/{N_\infty}) - c(t),
\ee
where $c(t)$ has a negligible value after the interval, $\tau$, between two timeline doses ($\tau=3$ weeks).
In other terms, chemoterapy effects start, periodically, at the beginning of each drug dose but almost completely decline after $\tau=3$ weeks and, therefore, the function $c(t)$ has a discontinuity on the days of treatment. 
By solving the previous eq.(2)  ( see appendix A)   for homogeneous, spherical symmetric configurations, the size reduction
after $n$ doses is given by
\be
ln({R_{n+1}}/{R_0})= ln({R_1}/{R_0}) \Sigma_{m=0}^n exp{(-m k \tau)},
\ee
where $R_{n+1}$ is the tumor radius after $n+1$ doses and, for each patient, the constant $k$ is determined by the initial cellularity (the second term in the growth law in eq.(1) is the fraction of duplicating tumor cells).

In the final result, (eq.3),  the function $c(t)$  does not explicitely appear: its contribution is hidden in the (measured) size after the first dose $R_1(\tau)$.
In this respect, the approach is independent on the model describing the chemotherapy effects.

\section{Validation, Results and Discussion}

\subsection{Validation: Patients and Therapy}
\vskip 10pt
{\bf Patients}
\vskip 10pt
This is a retrospective single centre study. Thirty-seven women, aged 36-78 years, with histologically proven operable breast cancer were evaluated. All tumours were tested for estrogen receptor (ER), progesterone receptor (PgR), HER 2 and ki 67 . Thirty-six patients showed positivity for ER (range $2-90\%$) and PgR (range $2-90\%$), HER2 $3+$ was present in  $5/37$ patients. Ki 67 was variable from $5$ to $30\%$. Median diameter of tumour , defined by imaging, was $43,5$ mm (range $21-72$ mm) .Four patients had clinical positivity for axillary nodes. Pregnant women  were excluded. ECOG-PS of all patients was 0 or 1. All patients had adequate haematological, renal and haepatic function. All patients had a normal left ventricular ejection fraction (LVEF > $50\%$).
\vskip 10pt
{\bf Treatment}
\vskip 10pt
Neoadjuvant chemotherapy corresponds to the use of a systemic treatment applied before locoregional treatment (surgery and /or radiotherapy) in order to obtain a more frequent conservating surgery, downgrading the tumour size. Major drugs used for breast cancer patients included anthracyclines and taxanes \cite{iv1} . Patients evaluated in our study received a median of five cycles (range 4-6) of every -3-week(q3w) ET (epirubicin 80 mg /m2 , paclitaxel 175 mg/m2 ) \cite{iv2,iv3}. Seven patients having HER2 $3+$ received integrate treatment with trastuzumab 6 mg/kg ( 8 mg/kg as loading dose). At the first follow up , after one chemotherapy administration , all patients had a tumour diameter reduction variable from $10$ to $70\%$. At the second follow up, after second chemotherapy administration , all patients showed a further diameter reduction included between $10$ and $30\%$. At the third follow up, $14/37$ patients continued to respond to treatment while the others showed a stable disease . At the fourth follow up , only one patient showed a futher tumour diameter reduction , the others continued to have a stabilization of disease and this was persisting at the remaining follow up \cite{iv4}.

\subsection{Results and Discussion}

The estimate of the tumor shrinkage in the dose sequence  follows immediately from eq.(3) and from the determination of $R_1(\tau)$. In Fig.1 and Fig.2 the numerical results are compared with data for  the second and the third treatment for all patients. The radius measurements had a $2-3\%$ statistical error and the error propagation has been taken into account. 

For the second dose, the ratio between predictions and data is within the prudential interval $1 \pm 0.1$ ($1$ indicates a perfect agreement) for 31 patients of the entire sample ($84\%$) 
and  the agreement is within $1\pm 0.2$ (see Fig.1) for $34/37$ ( $92\%$). 

For the third dose, in $29/37$  and $32/37$ cases the 
agreement is within the fiducial $1 \pm 0.1$, $1 \pm 0.2$  intervals (see Fig.2) respectively. 

The results in Fig.2 have been obtained by assuming an almost constant tumor size for stable disease.
On the other hand, one can ask if the diagnostic algorithm can give quantitative indications on  the stability of the  disease, i.e. if there is only a small reduction of the size after the treatment.

If a further reduction less than $10\%$ defines the stable disease condition, by applying the proposed algorithm, one gets that after the third treatment, $17/34$ patients continued to respond to the therapy (i.e. the tumor size decreases more than $10\%$)  to be compared with the clinical result of our sample, $14/37$ cases ( see Fig.3).

In Fig.4 analogous results are reported for the fourth dose, giving $4/37$ patients still responding  to the treatment ($1/37$ is the clinical result).

\begin{figure}
\epsfig{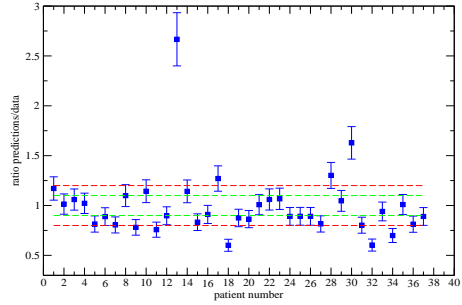}
\caption{Predictions to data ratio on the  tumor radius after the third dose. The green (red)  line indicates the $1 \pm 0.1$ ($1 \pm 0.2$) interval.}
\end{figure}

A homogeneous and spherical tumor is assumed in the previous sections, however these constraints can be easily removed and
the diagnostic algorithm can be improved in many directions, taking into account, for example, different geometrical tumor shapes or inhomogeneities.

The proposed approach can be applied in the clinical practice as follows.
Initially  the tumor size, $R_0$ and the cellularity are evaluated by usual methods and, after the first dose, one measures the  radius ( $R_1$). 
By those input data,  one then estimates the shrinkage of the tumor size due to the following treatments, i.e the values $R_2,R_3,..$, according to eq.(3).
If the tumor shrinkage observed after the second dose ( and before the third one)  turns out to be in agreement with  or larger than  the estimated results , $R_2$, then  the predictions should be considered clinically reliable. 
In particular, a forecast of stable disease ( defined by a reduction of the tumor radius less than $10\%$) suggests to stop the neo-adjuvant therapy and to proceed with surgery:  there is a clear signal that other drug doses are not effective to further reduce the  size.

A computational code can be easily implemented.

\section{Conclusions}

The good agreement between predictions and data for the treatment sequence suggests that the proposed method is a reliable starting point for a more quantitave description of neo-adjuvant chemotherapy effects and for an optimal management of patients, permitting to avoid unnecessary treatments and reducing economic costs. It should be further clarified that it has to be considered as a complementary tool to the standard chemotherapy response evaluation criteria in solid tumors \cite{iv1,viale,colleoni,CPSEG,recist1,recist2,choi}  and that it does not give any information on the overall survival probability, but a quantitative evaluation of the tumor size depletion.

\begin{figure}
\epsfig{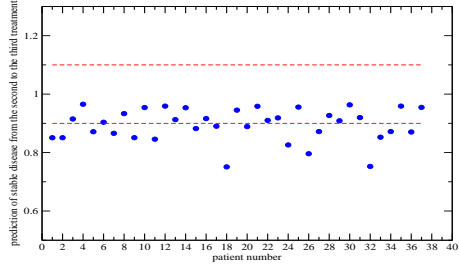}
\caption{Predictions of stabe disease after the third dose. The red  lines indicate the $1\pm 0.1$ interval.}
\end{figure}

\begin{figure}
\epsfig{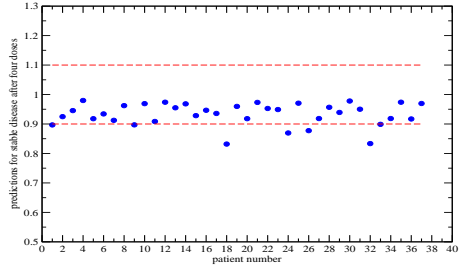}
\caption{Predictions of stabe disease after the fourth dose. The red  lines indicate the $1\pm 0.1$ interval.}
\end{figure}

\vskip 20pt

{\bf Acknowledgments}

The authors thank the Fondazione IOM. The work is partially supported by the project "DiOncoGen Diagnostica Innovativa", Codice CUP: G89J18000700007, Azione 1.1.5 del PO FESR SICILIA 2014/2020

\vfill
\eject

\vfill
\eject
\section{Appendix A}

The Gompertz law is solution of the equation  \cite{gompertz},
\be
\frac{1}{N} \frac{dN}{dt} = - k  ln (\frac{N}{N_\infty}),
\ee

which describes the macroscopic growth of a cancer cell population, without the effect of
chemotherapy \cite{norton1}. $N(t)$ is the cell number at time t, $k$ is a constant and $N_\infty$ is the maximum number of cells ( carrying capacity).

The drug treatment modifies the previous equation by inroducing the function $c(t)$ which decribes the chemotherapy effects during a single treatment, i.e. it is different from zero in  the interval, $\tau$, between two timeline doses ($\tau=3$ weeks). The general solution of the equation turns out to be
\be
y(t)= y(0)e^{-k (t-t_0)}-e^{-k (t-t_0)}\int_{t_0}^t c(t^{'})e^{+k (t'-t_0)} dt^{'}
\ee
where $y(t)=ln(N(t)/N_\infty)$, and $y(0)$ is the value at the initial time of the treatment $t_0$. By defining the initial time $t_0=0$ and
\be
I(\tau)=   e^{-k \tau} \int_0^\tau c(t^{'})   e^{+kt'}
\ee
the result of the first dose is given by
\be
y_1(\tau)= y(0)  e^{-k \tau} - I(\tau)
\ee
Since the function $c(t)$ is different from zero only for time interval $\tau$ after the beginning of any single dose,  the initial condition for the second treatment is $y_1(\tau)$ and at the end ($t=2\tau$) one gets
\be
y_2(2\tau)= y_1(\tau) e^{-k \tau}   - I(\tau)
\ee
By using the previous eq.(5) one  obtains
\be
y_2(\tau)- y(0) =  [y_1(\tau) - y(0)] [e^{-k \tau} + 1]
\ee
and, by iteration, at the end of $n+1$ treatments, one immediately gets 
\be
y_{n+1}[(n+1)\tau]-y(0)= [y_1(\tau) - y(0)] \Sigma_{m=0}^n e^{-m k \tau}
\ee
The previous equation written in term of tumor volume, for uniform density, is
\be
ln(\frac{V_{n+1}}{V_0})= ln(\frac{V_1}{V_0}) \Sigma_{m=0}^n e^{-m k \tau}
\ee
where $V_{n+1}$ is the tumor volume after $n+1$ doses. If one considers spherical symmetry, the previous equation can be written as
\be
ln(\frac{R_{n+1}}{R_0})= ln(\frac{R_1}{R_0}) \Sigma_{m=0}^n e^{-m k \tau}
\ee
where $R_{n+1}$ is the tumor radius after $n+1$ doses.

\end{document}